\documentclass[preprint,letter,preprintnumbers,nofootinbib,amsmath,amssymb,floatfix,superscriptaddress]{revtex4-1}

\usepackage{graphicx}
\usepackage{dcolumn}
\usepackage{bm}
\usepackage{amssymb}
\usepackage{epstopdf}

\begin{document}

\title{Parton energy loss in a classical strongly coupled QGP}
\author{Kevin Dusling and Ismail Zahed}
\affiliation{Department of Physics \& Astronomy, 
State University of New York, Stony Brook, NY 11794-3800, U.S.A.}
\date{\today}  

\begin{abstract}
We investigate the energy loss of heavy quarks in the gas, liquid 
and solid phase of a classical quark-gluon plasma (cQGP) using 
molecular dynamics simulations. The model consists of massive 
quarks and gluons interacting as a classical non-relativistic
colored Coulomb gas. We show that the electric force
decorrelates on a short time scale causing the energy loss
to be mostly diffusive and Langevin-like in the cQGP. We find
that the drag coefficient changes with the heavy quark mass, 
while the diffusion constant does not. The fractional collisional
energy loss is much larger than the leading order estimates from a 
wQGP (weakly coupled QGP) because of the core repulsion. Following recent suggestions,
we show how the cQGP results can be translated to the sQGP (strongly coupled QGP) results 
in the $T=(1-3)T_c$ range.
\end{abstract}

\maketitle  

\section{Introduction}

One possible signal for the formation of the quark-gluon plasma in heavy 
ion collisions is jet quenching.  Partons with a large transverse momentum 
are created from hard collisions between the partons of the nuclei involved 
in the initial heavy ion collision.  These produced high $p_T$ partons will 
traverse the collision region (possibly consisting of the quark-gluon plasma) 
within the first few fm/c.  Depending upon the properties of the medium 
traversed the parton will lose energy resulting in an experimentally observed 
jet quenching.

There have been numerous theoretical calculations of both collisional 
\cite{Djordjevic, Bjorken, TG, BT, Dumitru:2007rp, Schenke:2008gg, Thoma:2005uv, Thoma:2005aw} and radiative \cite{Jeon:2003gi,Baier:1996sk,Baier:1996kr,GVWZ,ASW} energy loss of 
partons in a QCD medium and it is seen that both effects play an important 
role \cite{Mustafa:2003vh,Mustafa,Wicks:2005gt} in the transverse momentum region where RHIC is most sensitive 
to jet quenching.   Even with these theoretical results at hand there is still no 
model which can explain the data \cite{Sh_book}, {\em i.e.} the observed 
quenching at RHIC energies is stronger then most theoretical predictions.

It should be mentioned that most of the energy loss results were computed in a 
weak coupling expansion which one would expect to converge as long as $\alpha_s \ll 1$.  
However, it is known from lattice results \cite{lat_alpha} that at the relevant 
temperatures probed at RHIC $\alpha_s$ reaches values of $\approx0.5$ and it should 
be checked how well the perturbative solution converges for the values of 
$\alpha_s$ probed in these experiments.

It is also known that the matter produced at RHIC cannot be weakly coupled 
but instead a good liquid.  The evidence for the sQGP \cite{EI1,Sh_sqgp} is 
large and growing but consists of the following two points: 1/ the observed 
collective flows at RHIC can be explained by hydrodynamics showing that the 
dissipative lengths are very short  2/  Binary bound states are seen to exist 
in lattice simulations above $T_c$ and are also predicted \cite{EI2} using 
lattice interparticle potentials.  We should point out to the reader that there has been arguments against bound states as well \cite{Koch:2005vg}.

Since perturbative methods generally fail in explaining strongly coupled systems 
other approaches have to be adopted.  For example, first principle calculations 
of the sQGP have been done using supersymmetric extensions of QCD via the 
AdS/CFT correspondence.  

The approach taken here, as was first discussed in \cite{MD1,MD2}, is to 
model the strongly interacting quark and gluon quasiparticles as a classical 
non-relativistic colored Coulomb gas.  This model is analyzed using Molecular 
Dynamics (MD) simulations in which real time correlators can be extracted.  
In \cite{MD1} decorrelation times, diffusion and viscosity was extracted for 
all phases.  It was found that when the results were extrapolated to the sQGP a 
diffusion constant of $D\approx0.1/T$ and viscosity to entropy density ratio 
of $\eta/s\approx 0.3$ was found.  In \cite{Young:2008he} Charmonium evolution in the sQGP was studied in a related framework. 

In this work we examine the energy loss of heavy quarks propagating through 
the sQGP. In section 2 we shortly review the key ingredients of the cQGP. We
show how the structure factor can be used to discriminate
the gas, liquid and solid phases. We also show that the electric force
decorrelates on a short time scale in the liquid and solid phase, meaning that
color probes become rapidly diffusive in the cQGP. In section 3, we numerically
assess the heavy quark diffusive properties and show that they are amenable
to a generic Langevin description. In the liquid phase, heavy quarks drag with
a drag coefficient that is smaller the larger the heavy quark mass. The diffusion
constant is independent of the heavy quark mass. In section 4, the relative 
energy loss of heavy quarks for the gas, liquid and solid phases are assessed. 
In section 5, we translate the cQGP results to the sQGP ones in the window of
temperatures $(1-3)T_c$. Our discussions and conclusions are in section 6.
Some useful units for comparison to the sQGP can be found in Appendix A. A
comparison of the energy loss with kinetic calculations is in Appendix B.

\section{Classical cQGP Model}

As mentioned in the introduction at temperatures close to $T_C$ quarks and gluons 
become quasiparticles with masses on the order of 3T.  We can model the sQGP as a system 
of massive non-relativistic particles interacting through longitudinal color electric fields.  
Magnetic effects are suppressed in the non-relativistic limit.  The specific Hamiltonian used 
in our model of the sQGP is 

\begin{align}
H=\sum_{\alpha i}\frac{p_{\alpha i}^2}{2m_\alpha}+\sum_{\alpha i\neq \beta j}
\left[  \frac{Q^a_{\alpha i}Q^b_{\beta j}}{|x_{\alpha i}-x_{\beta j}|}+V_{core} \right]
\label{HH}
\end{align}
where 

\begin{align}
V_{core}=\frac{1}{d}\left(\frac{1}{|x_{\alpha i}-x_{\beta j}|}\right)^d
\label{eq:core}
\end{align}
with $i,j=1..N_\alpha$ being a sum over all particles of specie 
$\alpha,\beta$= q,\={q},g having respective particle number $N_\alpha$.

The first term in the above Hamiltonian is the standard kinetic energy term.  
The second term is the colored coulomb interaction.  Again we have neglected 
any chromomagnetic interaction in the non-relativistic limit.  
Non-perturbative effects due to magnetic charges are discussed in \cite{Liao:2006ry, Liao:2007mj, Liao:2008jg, Liao:2008wb, Liao:2008vj, Liao:2008pu}.  
A short range repulsive potential was added by hand and is needed to give 
stability to the simulation and can be argued to mimic the effect of a quantum localization energy.  A more detailed study of the quantum corrections to potentials used in classical MD simulations can be found in \cite{Dusling:2007cn}.

The equations of motion can be derived from the usual Poisson brackets 
($\dot{O}_{\alpha i}=\{H,O_{\alpha i}\}$) where $O_{\alpha i}$ is the phase space coordinate 
of either position ($x_{\alpha i}$),  momentum ($p_{\alpha i}$) or color ($Q_{\alpha i}$) of 
particle $\alpha_i$.

The strength of the interparticle interaction is classified in the context of traditional 
electromagnetic plasmas, using the dimensionless parameter $\Gamma$, the ratio of the potential 
to kinetic energy:

\begin{align}
\Gamma=\frac{(Ze)^2}{a_{WS}T}
\end{align}
where $Ze, a_{WS}, T$ are respectively the ion charge, the Wigner-Seitz radius 
$a_{WS}=(3/4\pi n)^{1/3}$ and the temperature.  One usually defines the weakly coupled 
or gas regime for $\Gamma < 1$, a liquid regime for $\Gamma\approx1-10$ and a strongly 
coupled or solid regime for for $\Gamma > 10$.  

\begin{figure}
\includegraphics[scale=.85]{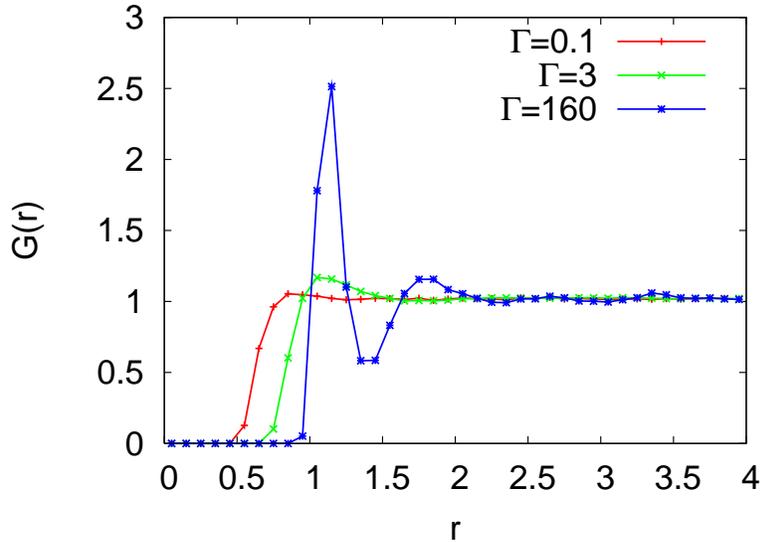}
\caption{(Color online) $G_d$ correlation function for 
$\Gamma\approx 0.1, 3.0, 160 \text{ and } t=0.$}
\label{fig:gdab}
\end{figure}

The gas, liquid and solid nature of the cQGP can be 
seen by looking at the structure factor for the different phases.  
In Fig.~\ref{fig:gdab} we show the distinct-correlation function $(G_d)$ defined as:

\begin{align}
G_d(\vec{x},t)=\frac{1}{N}<\sum^N_{i\neq j}\delta(\vec{x}+\vec{x}_i(0)-\vec{x}_j(t))>
\end{align}
for a gas, liquid and solid phase having $\Gamma\approx 0.1, 3, 160$ respectively.  
$G_d$ characterizes the probability to find two distinct particles having a separation 
$r$ at a time t.  One can see the that correlation among particles increases as one goes 
from the gas to liquid to glass phases.  It turns out, as discussed in appendix A, that 
we expect the properties of the sQGP to be consistent with the cQGP having $\Gamma\approx 3$.

\begin{figure}
\includegraphics[scale=.85]{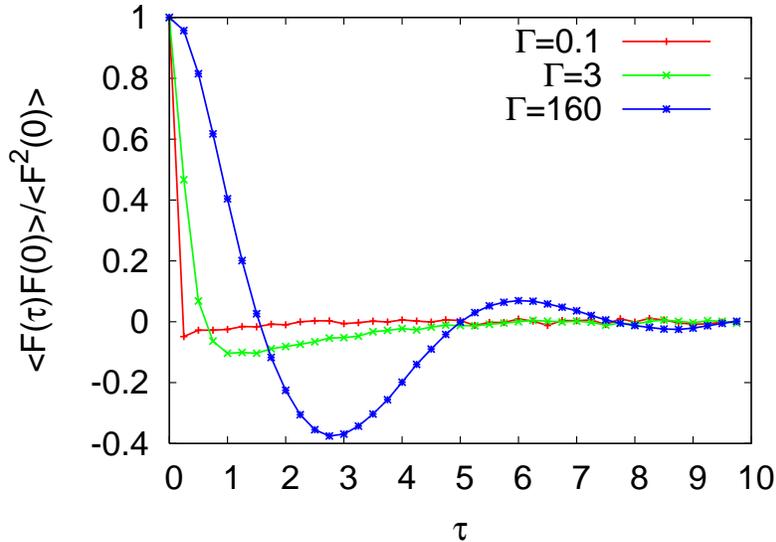}
\caption{Color electric force-force decorrelator for 
$\Gamma\approx 0.1, 3.0, 160$}
\label{fig:ff}
\end{figure}

The nature of the decorrelation times can be numerically investigated. The color
electric forces decorrelate on a short time scale in comparison to the velocity
decorrelation, meaning that color probes whether heavy or light are readily 
diffusive in the cQGP whether gas, liquid or solid. The electric force follows
from (\ref{HH}) as $Q_{i\alpha}\,F_{i\alpha}=-\partial H/\partial x_{i\alpha}$, and
the decorrelation function 

\begin{align}
G_{i\alpha}(t)= <F_{i\alpha} (t) F_{i\alpha} (0)>/<F_{i\alpha}(0)^2>
\end{align}
is shown in Fig.~\ref{fig:ff} for the gas, liquid and solid phase. The color
electric force decorrelates rapidly in the gas and liquid phase, and more
slowly in the ordered solid phase. In the liquid phase with $\Gamma=3$ the
decorrelation time is $t_F=\tau/2$ in simulation units. We translate these
units to physical units in Appendix A. In particular $t_F\approx 1/10T$ for the
sQGP.

\section{Diffusing Jets}

To analyze the evolution of jets in the cQGP at strong coupling, we will
use molecular dynamics simulations (see \cite{MD1} for more details) of 
the cQGP by integrating the above equations of motion of 64 particles confined 
to a box with periodic boundary conditions and mirror cubes on all sides.  
When equilibration of the system is reached an external particle is added 
to the simulation with a given momentum and mass (M) greater than the quark 
and gluon quasiparticle mass (m).  The evolution of this {\em probe} particle 
is tracked throughout the evolution of the system and it phase space coordinates 
are recorded.  This procedure is repeated until enough measurements have been made 
to make a statistically significant distribution of any phase space quantity.  For a given momentum bin we use $\approx 1500$ measurements.  Statistical errors are given by $\sqrt{N}$ of the number of runs. 

\begin{figure}[hbtp]
  \vspace{9pt}
  \centerline{\hbox{  
\includegraphics[scale=.75]{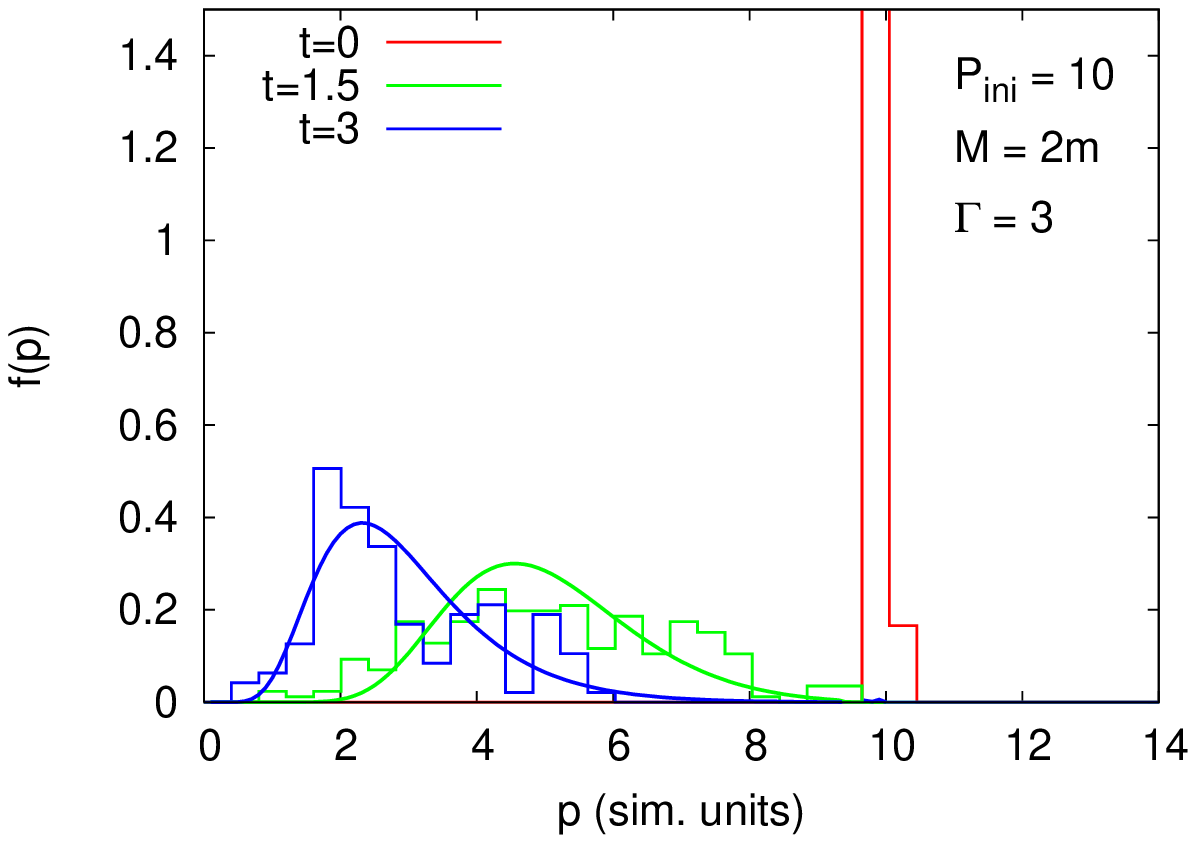}
\includegraphics[scale=.75]{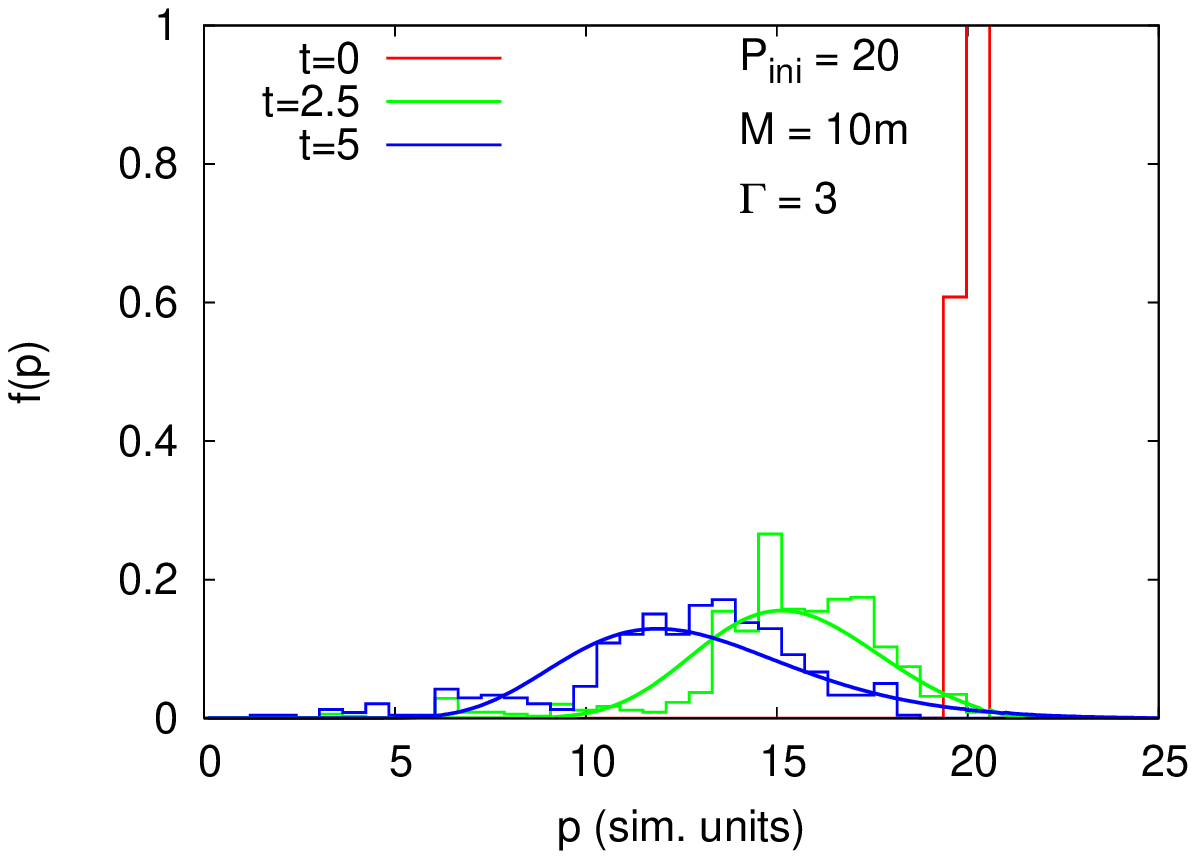}
    }
  }
  \vspace{9pt}
  \caption{ \label{fig:fp}(Color online) Momentum distribution function, f(p), shown for various values of time spent in 
   the liquid cQGP.  Left: $p_{init}=10 \text{ and } M=2m$, Right: $p_{init}=20 \text{ and } M=10m$.  
   All quantities are expressed in simulation units and measured from a simulation with $\Gamma=3$ corresponding to a temperature $T\approx 7$. }
 
\end{figure}

In Fig.~\ref{fig:fp} we show two examples of the evolution of the probe particle's 
momentum distribution function, f(p), as measured in the MD simulation in the
liquid phase with $\Gamma=3$. The first shows the evolution for a heavy quark mass 
with $M=2m$ and $p_{init}=10$ (large initial peak) and the second for a heavy quark 
mass with $M=10m$ and $p_{init}=20$ (large initial peak).  Each figure shows the 
initial momentum distribution and the distribution at two later times.  As the time 
of flight or distance increases a decrease in the average momentum of the probe as 
well as a broadening of the distribution function is seen. 

Since the color electric forces decorrelate promptly as shown in Fig.~\ref{fig:ff}, the massive jet
enters rapidly a diffusive regime. Assuming homogeneity in space, the diffusion is mostly
in momentum space. This is amenable to a non-relativistic Fokker-Planck equation 
(Ornstein-Uhlenbeck type) whereby the lowest two moments, i.e. drag and diffusion, are dominant. 
Restricting our analysis to the one-dimensional case for simplicity~\cite{Reif}

\begin{align}
\frac{\partial f}{\partial t}=\frac{\partial}{\partial p}\left[M_1(p) f\right]+
\frac{1}{2}\frac{\partial^2}{\partial p^2}\left[M_2(p) f\right]
\label{eq:fp}
\end{align}
where the lowest two moments $M_1$ and $M_2$ are related to the drag and diffusion
constant respectively,

\begin{align}
M_1(p)=\frac{\langle \delta p\rangle}{\delta t}\equiv -\eta(p)p \nonumber\\
M_2(p)=\frac{\langle (\delta p)^2 \rangle}{\delta t} \equiv \kappa_L(p)
\end{align}

\begin{figure}[hbtp]
  \vspace{9pt}
  \centerline{\hbox{  
\includegraphics[scale=.75]{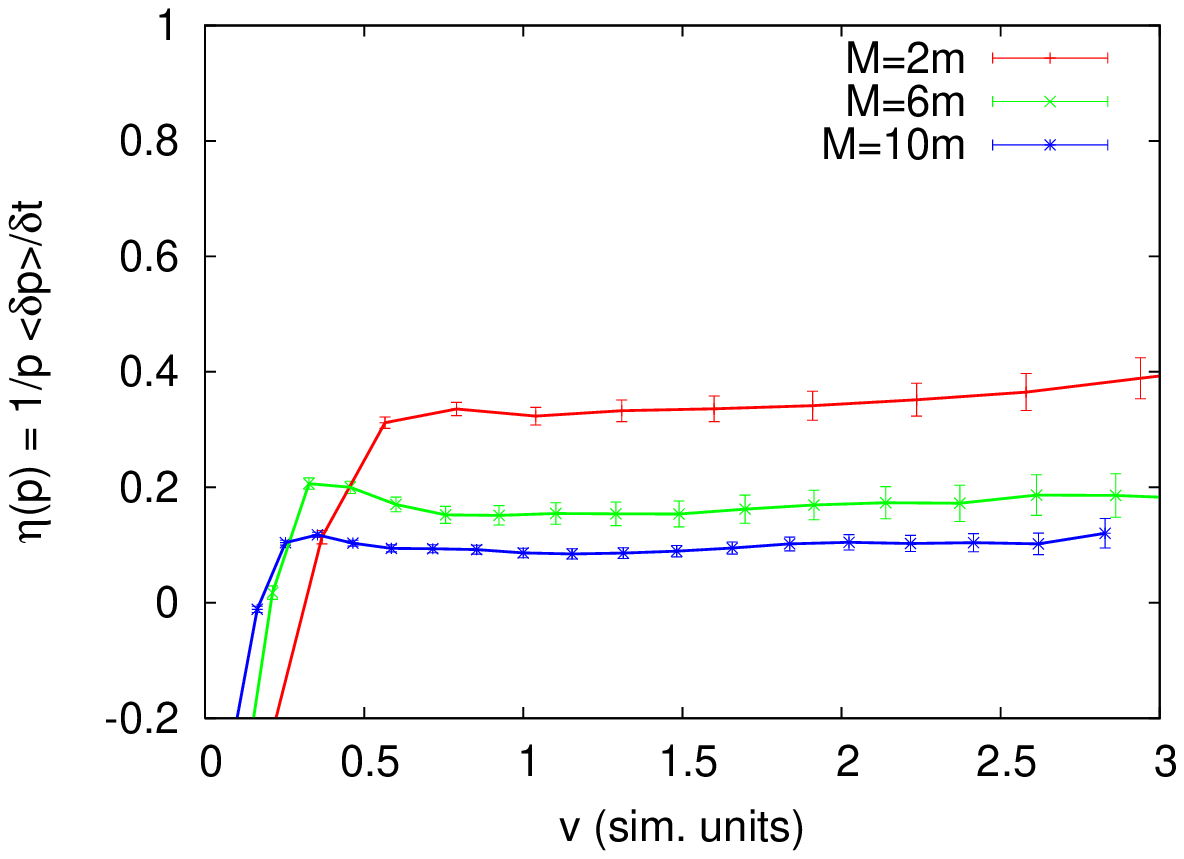}
\includegraphics[scale=.75]{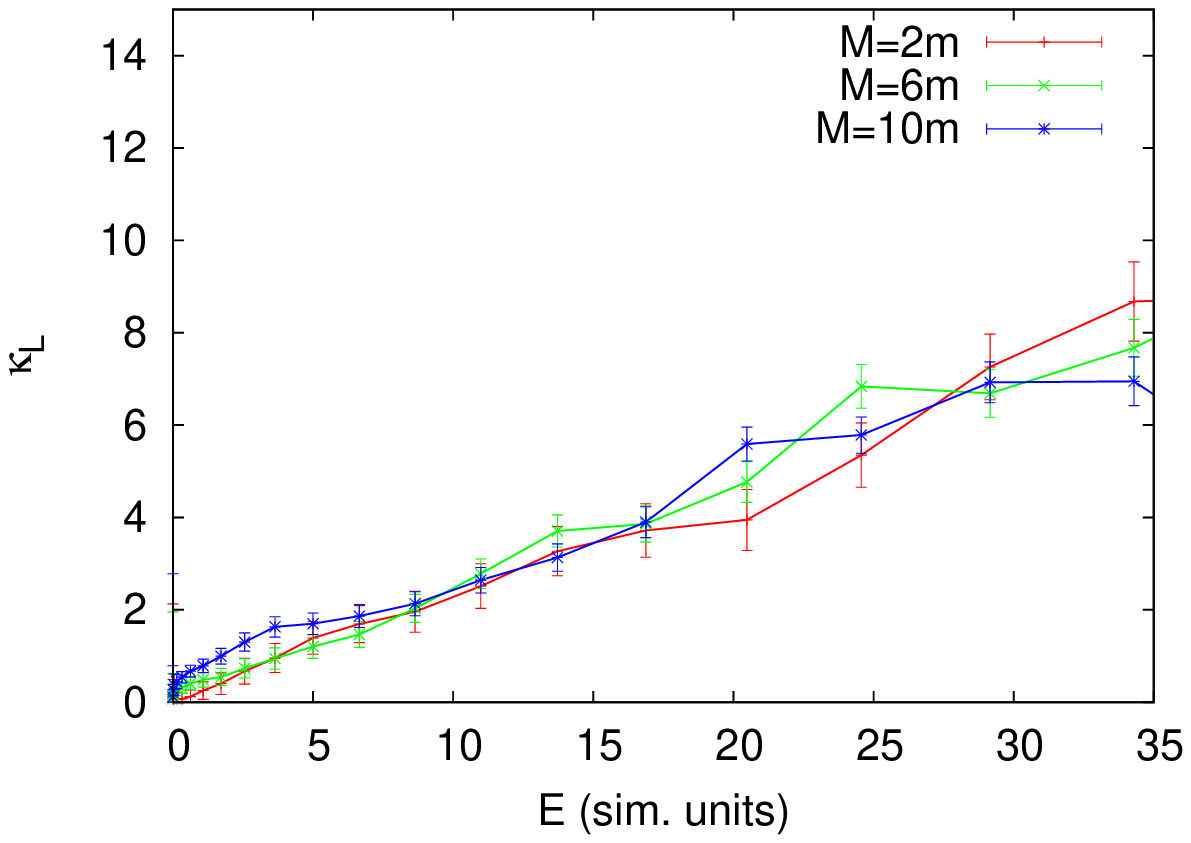}
    }
  }
  \vspace{9pt}
  \caption{(Color online) Measured transport coefficients $\eta$ and $\kappa_L$ for 
   three cases of the probe particle's mass.  All quantities are expressed in simulation 
   units and measured from a simulation with $\Gamma=3$.}
  \label{fig:dpdt}
\end{figure}

These two moments can be measured directly in our MD simulation.  The first moment, $M_1$, is 
related to the average momentum loss where $\eta(p)$ is the drag coefficient and as is seen in 
fig.~\ref{fig:dpdt} is independent of p for initial momentum greater than the thermal momentum 
of the system.  We therefore take $\eta(p)=0.35, 0.2 \text{ and } 0.1$ (in simulation units) 
for the various cases of the heavy quark mass having $M=2m, 6m, \text{ and } 10m$ respectively.  
Noticeably, the drag coefficient $\eta(p)$ decreases as the jet mass increases. 
The second moment, $M_2$, is related to the longitudinal momentum fluctuations ($\kappa_L$) and 
is $\propto E$ as shown in fig.~\ref{fig:dpdt}.  Within uncertainties $\kappa_L$ is independent 
of the heavy probe mass and we take $\kappa_L=0.25 E$.

Using the parameterizations of $\eta$ and $\kappa_L$ defined above the evolution equation \ref{eq:fp} 
can be solved numerically as shown in fig.~\ref{fig:fp} as solid curves.  Within the uncertainties of 
both the underlying model (soft collisions, one-dimensional) as well as our parameterization of 
transport coefficients the Fokker-Planck analysis shows good agreement with the MD distributions.

\section{Energy Loss}

From the momentum distribution function we can evaluate the mean energy of the external parton 
after traversing a distance L as:

\begin{align}
\langle E \rangle=\int_0^\infty{E\text{ }f(p,L)\text{ }dp}
\end{align}

and define the fractional energy loss as:

\begin{align}
\frac{\Delta E}{E} = \frac{E_0-\langle E \rangle}{E_0}
\end{align}

\begin{figure}
\centerline{\hbox{
\includegraphics[scale=.75]{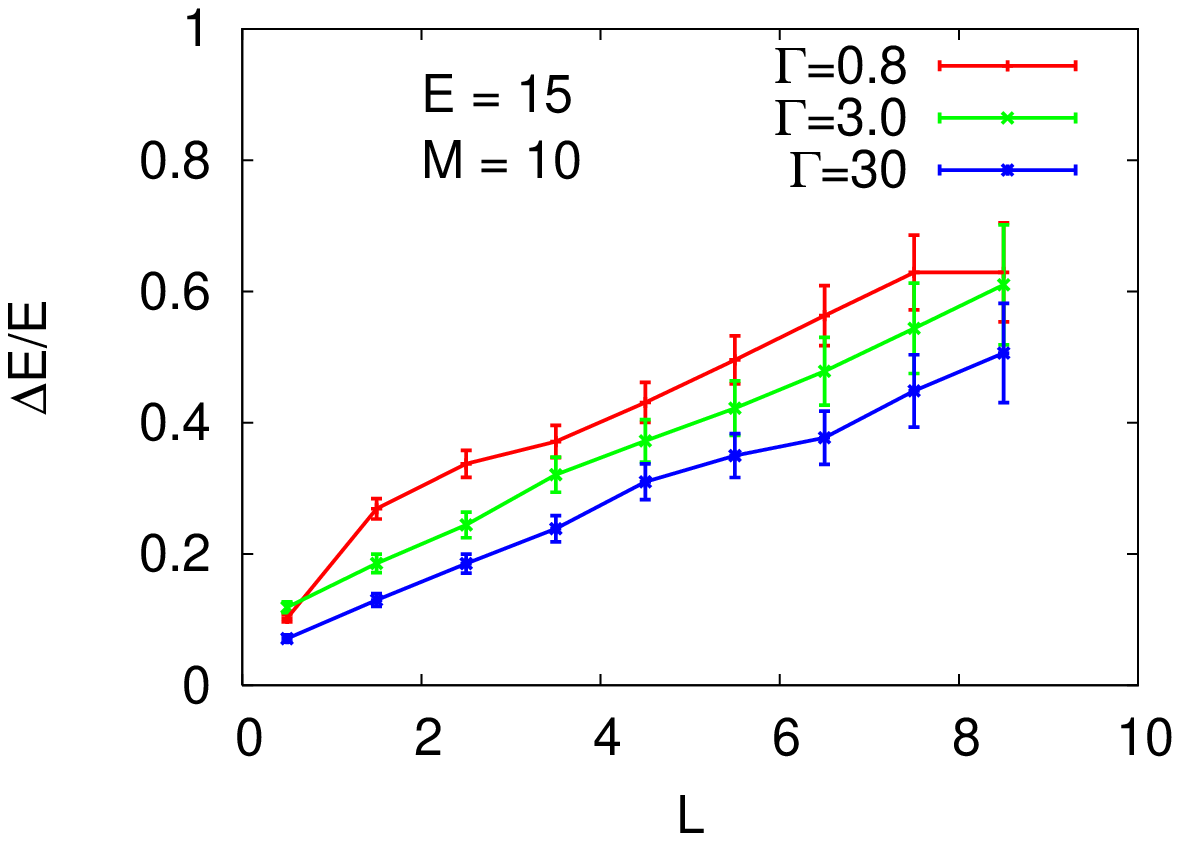}
\includegraphics[scale=.75]{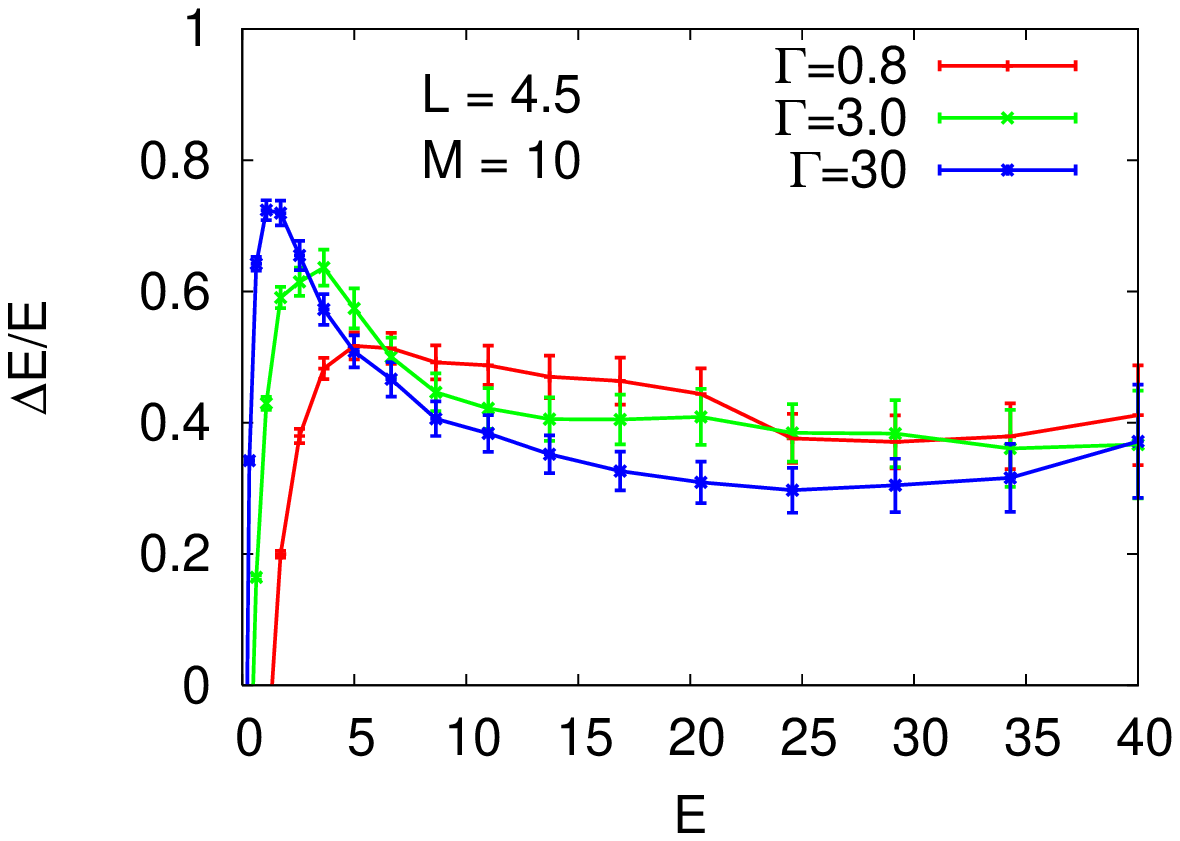}
}}
\caption{Fractional energy loss as a function of length and 
         energy measured from simulations having coupling parameters 
         of $\Gamma=0.8, 3.0, 30$. All quantities are expressed in simulation units.}
\label{fig:len}
\end{figure}

In figure~\ref{fig:len} we show the fractional energy loss as a function of length 
(in simulation units) for three different coupling parameters.  In all cases the external 
particle has a mass ten times that of the quasiparticle mass ({\em i.e.} $M=10$ in simulation units).  
For all three cases an approximately linear rise in energy loss as a function of length is seen 
which is expected for collisional loss only.  We also show the fractional energy loss as a function 
of initial energy (in simulation units) for three different values of coupling parameters.  
As expected, at thermal energies, the fractional energy loss is {\it negative} because a particle 
with $v=0$ can only gain momentum in collisions.  One sees that the fractional energy loss remains 
constant at high enough energies.  The fact that the energy loss is proportional to $E$ at high enough 
energy instead of increasing logarithmically as in the case for Coulomb collisions is due to the 
core potential (eq.~\ref{eq:core}).  Since the different coupling parameters $\Gamma$ are 
modified by changing the system temperature we don't expect to see differences in the energy loss 
when the probe's momentum is much greater than the thermal momentum $p\gg\sqrt{2MT}$ of the quasiparticles.

\section{Comparison to sQGP}

In order to compare the results from the MD simulation in the cQGP to the sQGP at RHIC
we follow the arguments of~\cite{MD1,MD2} as summarized in Appendix A. Specifically, 
we consider a plasma at a temperature $T=250$ MeV$\approx 1.5 T_C$ corresponding to a 
coupling parameter $\Gamma=3$.  The unit of mass is rescaled using $[m]=[3T]\approx0.75$ GeV, 
so a heavy probe with mass 2, 6 and 10 times the quasiparticle mass as used in the simulation 
corresponds to masses of 1.5 GeV, 4.5 GeV and 7.5 GeV respectively in the sQGP.  
As mentioned earlier the plasma consists of quasiparticles with $m\approx3T\approx0.75$ GeV 
at $T=1.5T_C$.  The drag coefficient has units $[\eta]=[1/\tau]=[5.1T]\approx6.5$ fm$^{-1}$.  
Assuming that the drag is independent of $p$ as we showed for a large range of momenta 
we find that $\eta=2.3, 1.3, 0.65$ fm$^{-1}$ for probe masses of $M=1.5, 4.5, 7.5$ GeV respectively.

\begin{figure}[hbtp]
\includegraphics[scale=.75]{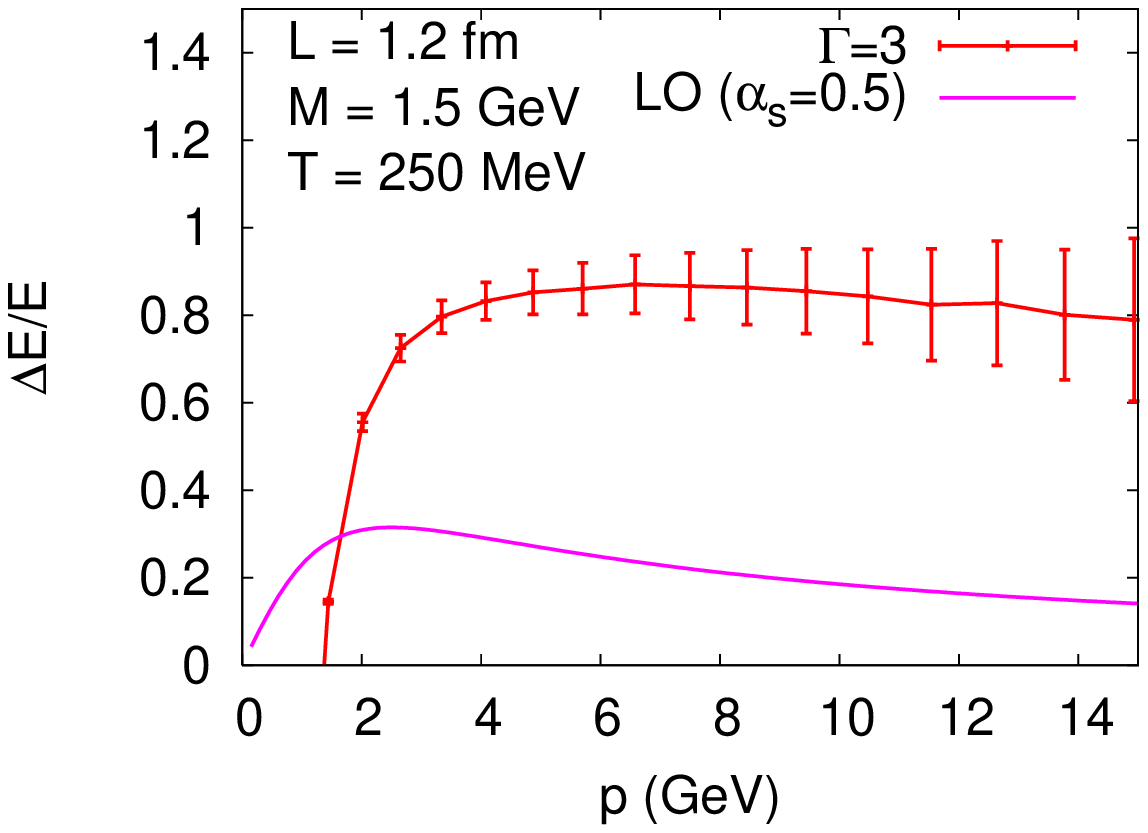}
\vspace{9pt}
\includegraphics[scale=.75]{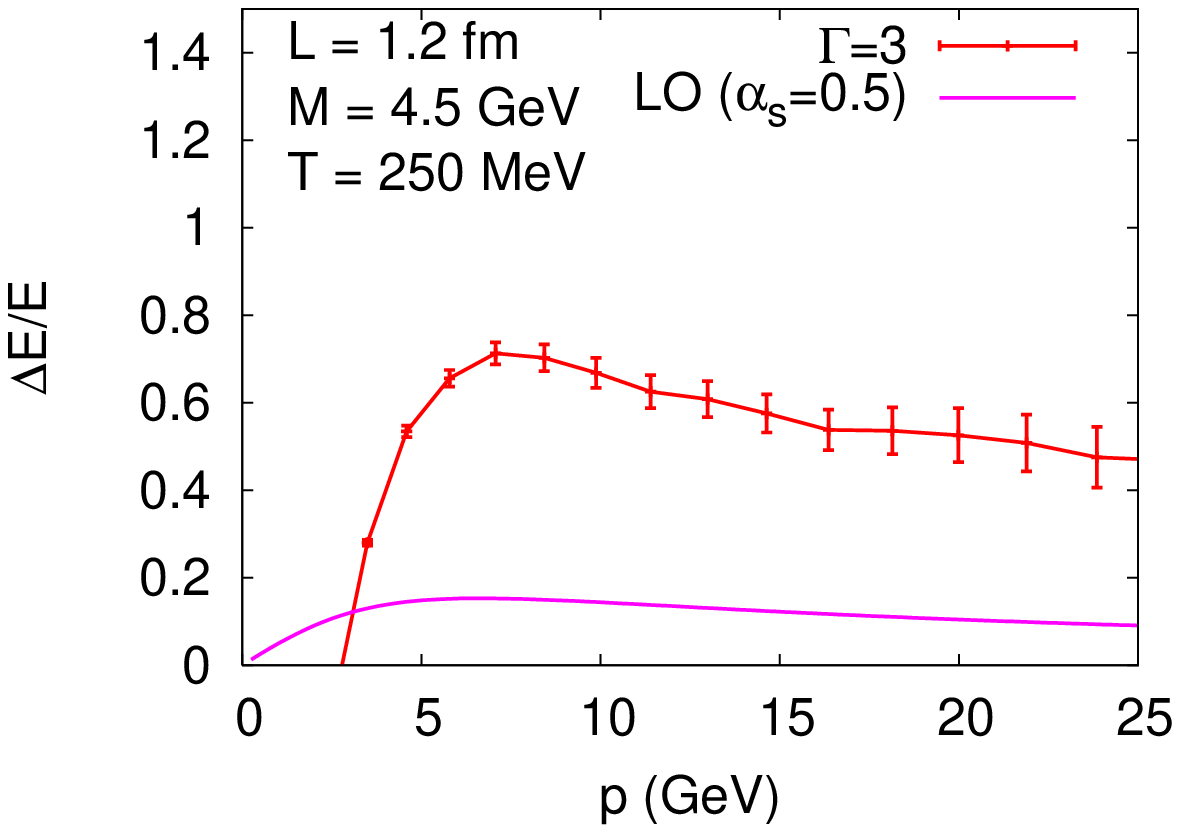}
\vspace{9pt}
\includegraphics[scale=.75]{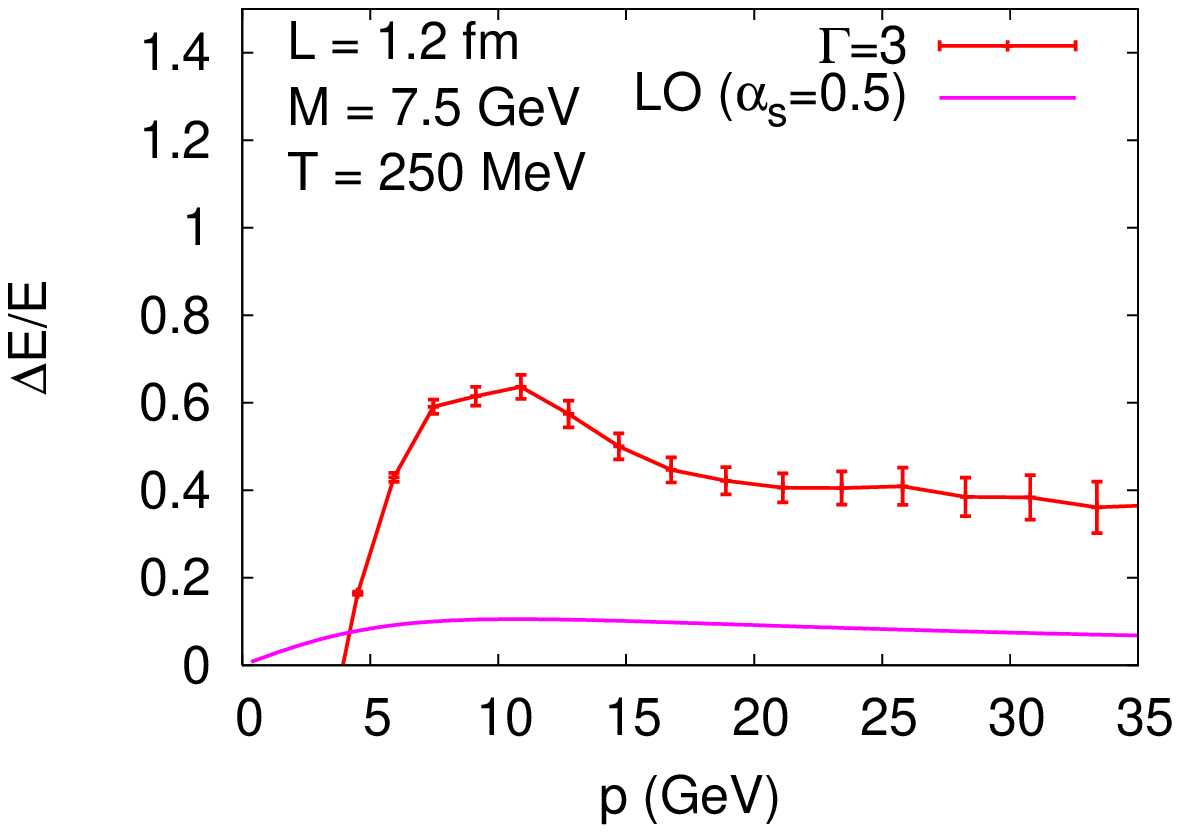}
  \vspace{9pt}
  \caption{ Fractional collisional energy loss as a function of momentum for 
   charm, bottom and heavy quarks (top, middle and bottom figures respectively) in the sQGP.}
  \label{fig:eloss_avg_unit}
\end{figure}

It is also useful to look at the energy loss after the probe particle travels a finite 
length L (here taken to be about the length of our simulation region or 4.5 simulation units) 
which corresponds to $1.2$ fm in the sQGP.  In fig.~\ref{fig:eloss_avg_unit} we show the 
fractional energy loss as a function of the probe particle's initial momentum as calculated 
from the MD simulation (points with error bars).  For comparison we also show the leading order in $\alpha_s$ collisional energy loss of the 
scattering of heavy quarks off of massless quarks and gluons \cite{MLB,BT}.  The sensitivity to the schematic form of the core potential is smaller at larger values of momentum.  This is verified by a comparison of the MD simulation results to analytic calculations preformed in \cite{Cho:2009ze}.

\begin{align}
\frac{dE}{dx}=-\frac{8\pi\alpha_s^2 T^2}{3v}(1+\frac{n_f}{6})
(1-\frac{1-v^2}{2v}\ln{\frac{1+v}{1-v}})\ln{\frac{q^{max}}{q^{min}}}
\label{eq:dedx}
\end{align}
where the lower cutoff of the momentum transfer is taken as the Debye mass: 
$q^{min}\approx2T$ and the upper cutoff $q^{max}\approx\sqrt{4TE}$ is taken 
from \cite{Bjorken}.   In order to compare the above equation to our simulation results 
we take $\Delta E/E \approx L/E\cdot (-dE/dx)$ where $L=1.2 fm$.  

\section{Discussion and Conclusions}

First, it should be mentioned, that the energy loss in equation \ref{eq:dedx} acts as 
an upper limit for other energy loss calculations as presented in \cite{BT,Djordjevic}.  
Comparing the MD simulation results to the leading order results in eq.~\ref{eq:dedx} one 
immediately sees a much larger fractional energy loss for the sQGP. As much as a factor 
of five in the high momentum region.  The main physical differences that account for the 
large increase are: 1/ the use of quasiparticle quark and gluon masses ($m\approx3T$) much 
greater than the current quark masses 2/ the use of the hard core potential to {\em mock up} 
the effects of quantum repulsion at short distances. 3/ any non-local many body interactions 
induced by the coulomb and core potential are resummed to all orders using MD.

One should note that if the current lattice results at $T\approx1.5T_C$ are modified the 
same MD simulation results can be used by simply applying a different rescaling of the units 
consistent with the data.  Since the fractional energy loss is mostly constant as a function 
of momentum, a rescaling of the momentum will not change the results in the high energy region.  
However, both the length of the medium and the mass of the heavy probe particle will need to be 
adjusted accordingly.  The change in length unit will also modify the result for the drag coefficient.

In conclusion we have computed the parton energy loss in a strongly coupled quark-gluon plasma 
using a classical molecular dynamic simulation.  The model consists of massive strongly interacting 
quark and gluon quasiparticles at $T=(1-3)T_C$ interacting as a classic non-relativistic colored 
Coulomb gas.  We find that the fractional collisional energy loss is much larger compared to the 
leading order estimates in a wQGP.
\\

{\bf Acknowledgments.}
\\
This work was supported in part by US-DOE grants DE-FG02-88ER40388 and DE-FG03-97ER4014.
KD thanks INT (Seattle) for its hospitality and the Department of Energy for the support 
during the Workshop INT--06--3.

\appendix
\section{Units and Comparison to sQGP}

The equations of motion provided above are integrated over time given some initial configuration 
of phase space coordinates.  It is convenient to run the evolution in simulation units which we now 
discuss following~\cite{MD1,MD2}. First, the unit of length is set by the minimum of the potential 
which has the form:

\begin{align}
V=\frac{g^2}{\lambda}\left[ Q\cdot Q\frac{\lambda}{r}+ \frac{1}{d}(\frac{\lambda}{r})^d \right]
\end{align}
where $\lambda=r_{min}$ (for any value of d) sets the basic length scale in which all distances 
are measured.  The time unit is set by the plasma frequency of the system:

\begin{align}
\tau=\omega^{-1}_p=\left(\frac{m}{4\pi n e^2}\right)^{1/2}
\end{align}
and the unit of mass is defined as the particle mass.  For example, in a simulation of 
only one particle specie all masses are equal to one.

In order to extract results from the cQGP simulation about the physical sQGP, all that 
is required is a re-scaling of the three basic units of length, time and mass.  For now this is 
done at a temperature of $T=1.5-3 T_c$ where most lattice data is available.
The unit of mass is set by the mass of the quark and gluon quasiparticles taken from 
lattice data: $m\approx3T$ at $T=1.5T_c$ as discussed earlier.

The effective interparticle potential is given by:

\begin{align}
V_{eff}=\frac{\hbar^2}{2mr^2}-\frac{C\alpha_s}{r}
\end{align}
where C is the pertinent Casimir for quarks and gluons.  The length unit in the cQGP is 
set by the minimum of the potential, $\lambda = {\hbar^2}/{mC\alpha_s}$.
With $\alpha_s\approx0.5$ at these distances and averaging over the color casimir for 
quarks and gluons assuming that all three species ($g, \bar{q}, q$) are equally 
represented, we have $<\alpha_s C>\approx1$ within uncertainties of the model.
This leads to $\lambda\approx\frac{1}{3T}$ in units where $\hbar=c=1$.

Finally the time unit is given by $\left( \frac{4\pi n <\alpha_s C>}{m}\right)^{-1}$.  
The density of quasiparticles (n) is estimated as the density of black body radiation 
photons multiplied by the effective degrees of freedom: $n\approx(0.244T^3)(8+6N_f)\approx6.3T^3$.
Then the time unit is measured in $\tau\approx\frac{1}{5.1T}$.

\section{Comparison with Kinetics}

\begin{figure}
\includegraphics[scale=.65]{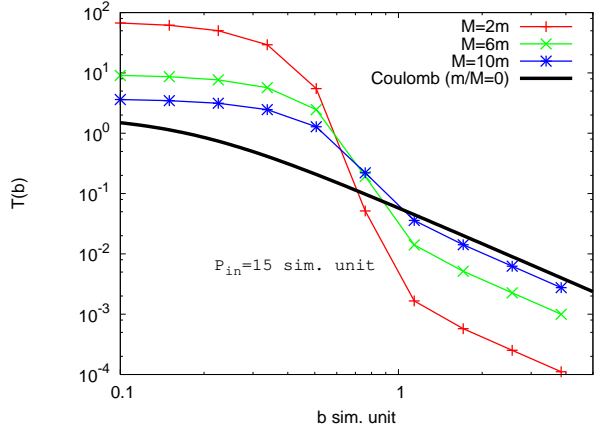}
\caption{Energy transfer as a function of impact parameter for a heavy quark 
         with $P_{in}=15$ and mass M=2m, 6m, and 10m}
\label{fig:Tb}
\end{figure}

In order to understand the collisional energy loss better we compare the results 
from the full MD simulation with a simple kinetic calculation.  We assume that the 
heavy quark undergoes two body scattering with a plasma of massive quasiparticles 
of density n.  When the heavy particle passes through the plasma it sees 
quasiparticles at all possible impact parameters.  The energy loss is given by~\cite{Jackson}:

\begin{align}
\label{eq:dEdx}
\frac{dE}{dx}=2\pi n\int_0^{b_{max}} T(b) b db
\end{align}
where $T(b)$ is the energy transfer to the heavy particle in a two body collision 
with a quasiparticle at rest at an impact parameter $b$.  Fig.~\ref{fig:Tb} shows the 
energy transfer as a function of impact parameter for three heavy quark masses and an 
initial momentum of $P_{in}=15$ in simulation units.  The results were calculated for 
the same coulomb+hard core potential as used in the MD simulations.  For comparison 
the solid curve shows the analytic result for scattering from a coulomb potential in 
the limit that $m/M\to 0$:  

\begin{align}
T(b)=\frac{2\alpha^2}{E}\frac{1}{(\frac{\alpha}{2E})^2+b^2}
\end{align}
For distances larger than one $T(b)$ increases towards the Coulomb case as the mass 
is increased as is expected.  For distances smaller then one there is a large 
enhancement in the energy transfer due to the strong repulsive core of the MD potential.

\begin{figure}
\includegraphics[scale=.65]{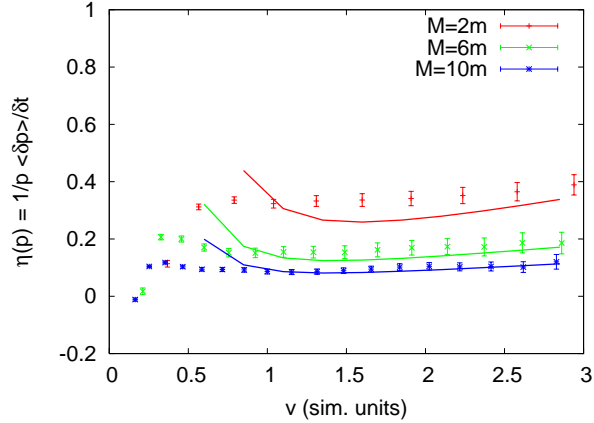}
\caption{Drag coefficient as a function of heavy quark velocity.  The data points are 
         the results of the MD simulation.  The solid curves are the results of 
          eq.~\ref{eq:dEdx} for velocities greater then the thermal 
           velocity ($v_{th}\approx \sqrt{\frac{2T}{M}}$).}
\label{fig:compare}
\end{figure}

In fig.~\ref{fig:compare} we show the drag given by equation~\ref{eq:dEdx} compared to 
the MD simulation results already presented from fig.~\ref{fig:dpdt}.   The comparison is 
only shown for velocities larger then the thermal velocity where equation 
\ref{eq:dEdx} holds.  There is good agreement between this simple model and the full MD results.

\begin{thebibliography}{MM}

\bibitem{Djordjevic} Magdalena Djordjevic, arXiv:nucl-th/0603066
\bibitem{Bjorken} J. D. Bjorken, FERMILAB-PUB-82-059-THY (unpublished)
\bibitem{TG} M. H. Thoma and M. Gyulassy, Nucl, Phys. B {\bf 351}, 491 (1991).
\bibitem{BT} E. Braaten and M. H. Thoma, Phys. Rev. D {\bf 44}, 2625 (1991).
\bibitem{Dumitru:2007rp}
  A.~Dumitru, Y.~Nara, B.~Schenke and M.~Strickland,
  Phys.\ Rev.\  C {\bf 78}, 024909 (2008)
  [arXiv:0710.1223 [hep-ph]].
\bibitem{Schenke:2008gg}
  B.~Schenke, M.~Strickland, A.~Dumitru, Y.~Nara and C.~Greiner,
  Phys.\ Rev.\  C {\bf 79}, 034903 (2009)
  [arXiv:0810.1314 [hep-ph]].
\bibitem{Thoma:2005uv}
  M.~H.~Thoma,
  J.\ Phys.\ G {\bf 31}, L7 (2005)
  [arXiv:hep-ph/0503154].
\bibitem{Thoma:2005aw}
  M.~H.~Thoma,
  Nucl.\ Phys.\  A {\bf 774}, 307 (2006)
  [arXiv:hep-ph/0509154].
\bibitem{Jeon:2003gi}
   S.~Jeon and G.~D.~Moore,
   Phys.\ Rev.\  C {\bf 71}, 034901 (2005)
 \bibitem{Baier:1996sk}
   R.~Baier, Y.~L.~Dokshitzer, A.~H.~Mueller, S.~Peigne and D.~Schiff,
   Nucl.\ Phys.\  B {\bf 484}, 265 (1997)
   [arXiv:hep-ph/9608322].
 \bibitem{Baier:1996kr}
   R.~Baier, Y.~L.~Dokshitzer, A.~H.~Mueller, S.~Peigne and D.~Schiff,
   Nucl.\ Phys.\  B {\bf 483}, 291 (1997)
   [arXiv:hep-ph/9607355].
\bibitem{GVWZ} M. Gyulassy, I. Vitev, X. N. Wang and B. W. Zhang, 
Quark Gluon Plasma 3, editors: R. C. Hwa and X. N. Wang, World Scientific, Singapore, 123 (2003) (nucl-th/0302077).
\bibitem{ASW}
1483 B. G. Zakharov, JETP Lett. {\bf 63}, 952 (1996), hep-ph/9607440.  B. G. Zakharov, JETP Lett. {\     bf 65}, 615 (1997), hep-ph/9704255.  B. G. Zakharov, Phys. Atom. Nucl. {\bf 61}, 838 (1998), he     p-ph/9807540.  C. A. Salgado and U. A. Wiedemann, Phys. Rev. {\bf D68}, 014008 (2003), hep-ph/0     302184.  U. A. Wiedemann, Nucl. Phys. {\bf B582}, 409 (2000), hep-ph/0003021.  U. A. Wiedemann,      Nucl. Phys. {\bf B588}, 303 (2000), hep-ph/0005129.  C. A. Salgado and U. A. Wiedemann, Phys.      Rev. Lett.  {\ bf 89}, 092303 (2002), hep-ph/0204221. N. Armesto, C. A. Salgado, and U. A.Wiede     mann, Phys. Rev. Lett. {\bf 94}, 022002 (2005), hep-ph/0407018.
\bibitem{Mustafa:2003vh}
  M.~G.~Mustafa and M.~H.~Thoma,
  Acta Phys.\ Hung.\  A {\bf 22}, 93 (2005)
  [arXiv:hep-ph/0311168].
\bibitem{Mustafa} M. G. Mustafa, Phys. Rev. C {\bf 72}, 014905 (2005).
\bibitem{Wicks:2005gt}
  S.~Wicks, W.~Horowitz, M.~Djordjevic and M.~Gyulassy,
  Nucl.\ Phys.\  A {\bf 784}, 426 (2007)
  [arXiv:nucl-th/0512076].
\bibitem{MLB} M. Le Bellac, {\em Thermal Field Theory} (Cambridge University Press, 1996).
\bibitem{Cho:2009ze}
  S.~Cho and I.~Zahed,
  arXiv:0910.1548 [nucl-th].

\bibitem{lat_alpha} O. Kaczmarek , F. Karsch, F. Zantow and P. Petreczky, Phys. Rev. D {\bf 70} (2004) 074505.

\bibitem{Ducati} M. B. Gay Ducati, V. P. Gon\c{c}alves, L. F. Mackendanz, arXiv:hep=ph/0506241
\bibitem{Sh_book} E. V. Shuryak, The QCD Vacuum, Hadrons and Superdense Matter Second Edition, World Scientific, Singapore, 497 (2004). 

\bibitem{Sh_sqgp} E. V. Shuryak, arXiv:hep-ph/0608177.

\bibitem{EI1} Edward V. Shuryak and Ismail Zahed, Phys. Rev. C {\bf 70}, 021901 (2004).
\bibitem{EI2} Edward V. Shuryak and Ismail Zahed, Phys. Rev. D {\bf 70}, 054507 (2004).
\bibitem{Koch:2005vg}
  V.~Koch, A.~Majumder and J.~Randrup,
  Phys.\ Rev.\ Lett.\  {\bf 95}, 182301 (2005)
  [arXiv:nucl-th/0505052].
\bibitem{MD1}Boris A. Gelman, Edward V. Shuryak and Ismail Zahed, Phys. Rev. C {\bf 74} 044908 (2006).
\bibitem{Young:2008he}
  C.~Young and E.~Shuryak,
  Phys.\ Rev.\  C {\bf 79}, 034907 (2009)
  [arXiv:0803.2866 [nucl-th]].
\bibitem{MD2}Boris A. Gelman, Edward V. Shuryak and Ismail Zahed, Phys. Rev. C {\bf 74} 044909 (2006).
\bibitem{Liao:2006ry}
  J.~Liao and E.~Shuryak,
  Phys.\ Rev.\  C {\bf 75}, 054907 (2007)
  [arXiv:hep-ph/0611131].
\bibitem{Liao:2007mj}
  J.~Liao and E.~Shuryak,
  Phys.\ Rev.\  C {\bf 77}, 064905 (2008)
  [arXiv:0706.4465 [hep-ph]].
\bibitem{Liao:2008jg}
  J.~Liao and E.~Shuryak,
  Phys.\ Rev.\ Lett.\  {\bf 101}, 162302 (2008)
  [arXiv:0804.0255 [hep-ph]].
\bibitem{Liao:2008wb}
  J.~Liao and E.~Shuryak,
  J.\ Phys.\ G {\bf 35}, 104058 (2008)
  [arXiv:0804.3102 [hep-ph]].
\bibitem{Liao:2008vj}
  J.~Liao and E.~Shuryak,
  arXiv:0804.4890 [hep-ph].
\bibitem{Liao:2008pu}
  J.~Liao and E.~Shuryak,
  arXiv:0809.2419 [hep-ph].


\bibitem{Dusling:2007cn}
  K.~Dusling and C.~Young,
  arXiv:0707.2068 [nucl-th].

\bibitem{Hartmann} P. Hartmann, Z. Donko, P. Levai and G. J. Kalman, Nucl. Phys. A {\bf 774}, 881 (2006) [arXiv:nucl-th/0601017]

\bibitem{Reif} F. Reif, Fundamentals of Statistical Physics, Sect. 15, McGraw-Hill (1965).

\bibitem{MT} S. Mr\'{o}wczy\'{n}ski and M Thoma, nucl-th/0701002

\bibitem{Jackson} J. D. Jackson, "Classical Electrodynamics, Chapt. 13, John Wiley \& Sons, Inc.  (1999).
\end{thebibliography}
\end{document}